\documentstyle[12pt]{article}

\newtheorem{theorem}{Theorem}
\newcommand{\proof}{{\bf{Proof. }}}

\begin{document}

\author{C. Bizdadea\thanks{e-mail address: bizdadea@central.ucv.ro},
A. Constantin, S. O. Saliu\thanks{e-mail address: osaliu@central.ucv.ro}\\
Department of Physics, University of Craiova\\
13 A. I. Cuza Str., Craiova RO-1100, Romania}
\title{Note on irreducible approach to reducible second-class constraints}

\maketitle

\begin{abstract}
An irreducible canonical approach to reducible second-class constraints is
given. The procedure is illustrated on gauge-fixed two-forms.

\end{abstract}

The canonical approach to systems with reducible second-class constraints is
quite intricate, demanding a modification of the usual rules as the matrix
of the Poisson brackets among the constraints is not invertible. Thus, it is
necessary to isolate a set of independent constraints, and then construct
the Dirac bracket \cite{1,2} with respect to this set. The split of
the constraints may lead to the loss of important symmetries, so it should
be avoided. As shown in \cite{3,4,5,6,7}, it is however possible to
construct the Dirac bracket in terms of a noninvertible matrix without
separating the independent constraint functions. A third possibility is to
substitute the reducible second-class constraints by some irreducible ones
and further work with the Dirac bracket based on the irreducible
constraints. This idea has been suggested in \cite{8} mainly in the context
of two- and three-form gauge fields.

Although the idea based on irreducible second-class constraints is known, a
general irreducible procedure equivalent to the reducible one has not been
developed so far. This is the aim of this letter.

We start with a system locally described by $N$ canonical pairs $%
z^{a}=\left( q^{i},p_{i}\right) $, subject to the second-class constraints 
\begin{equation}
\chi _{\alpha _{0}}\left( z^{a}\right) \approx 0,\;\alpha _{0}=1,\cdots
,M_{0}.  \label{1}
\end{equation}
For simplicity, we take all the phase-space variables to be bosonic. In
addition, we presume that the functions $\chi _{\alpha _{0}}$ are not
independent, there existing some nonvanishing functions $Z_{\;\;\alpha
_{1}}^{\alpha _{0}}$ such that 
\begin{equation}
Z_{\;\;\alpha _{1}}^{\alpha _{0}}\chi _{\alpha _{0}}=0,\;\alpha
_{1}=1,\cdots ,M_{1}.  \label{2}
\end{equation}
Moreover, we assume that $Z_{\;\;\alpha _{1}}^{\alpha _{0}}$ are independent
and (\ref{2}) are the only reducibility relations with respect to the
constraints (\ref{1}). These constraints are purely second-class if any
maximal, independent set of $M_{0}-M_{1}$ constraint functions $\chi _{A}$ ($%
A=1,\cdots ,M_{0}-M_{1}$) among the $\chi _{\alpha _{0}}$ is such that the
matrix 
\begin{equation}
C_{AB}=\left[ \chi _{A},\chi _{B}\right] ,  \label{3}
\end{equation}
is invertible. In terms of independent constraints, the Dirac bracket takes
the form 
\begin{equation}
\left[ F,G\right] ^{*}=\left[ F,G\right] -\left[ F,\chi _{A}\right]
M^{AB}\left[ \chi _{B},G\right] ,  \label{4}
\end{equation}
where $M^{AB}C_{BC}\approx \delta _{\;\;C}^{A}$. We can rewrite the Dirac
bracket in (\ref{4}) without finding a definite subset of independent
second-class constraints as follows. We start with the matrix 
\begin{equation}
C_{\alpha _{0}\beta _{0}}=\left[ \chi _{\alpha _{0}},\chi _{\beta
_{0}}\right] ,  \label{5}
\end{equation}
that is not invertible because 
\begin{equation}
Z_{\;\;\alpha _{1}}^{\alpha _{0}}C_{\alpha _{0}\beta _{0}}\approx 0.
\label{6}
\end{equation}
If $d_{\;\;\alpha _{0}}^{\alpha _{1}}$ is solution to the equation 
\begin{equation}
d_{\;\;\alpha _{0}}^{\alpha _{1}}Z_{\;\;\beta _{1}}^{\alpha _{0}}\approx
\delta _{\;\;\beta _{1}}^{\alpha _{1}},  \label{7}
\end{equation}
then we can introduce a matrix \cite{6} $M^{\alpha _{0}\beta _{0}}$ through
the relation 
\begin{equation}
M^{\alpha _{0}\beta _{0}}C_{\beta _{0}\gamma _{0}}\approx \delta
_{\;\;\gamma _{0}}^{\alpha _{0}}-Z_{\;\;\alpha _{1}}^{\alpha
_{0}}d_{\;\;\gamma _{0}}^{\alpha _{1}},  \label{8}
\end{equation}
with $M^{\alpha _{0}\beta _{0}}=-M^{\beta _{0}\alpha _{0}}$. Then, the
formula \cite{6} 
\begin{equation}
\left[ F,G\right] ^{*}=\left[ F,G\right] -\left[ F,\chi _{\alpha
_{0}}\right] M^{\alpha _{0}\beta _{0}}\left[ \chi _{\beta _{0}},G\right] ,
\label{10}
\end{equation}
defines the same Dirac bracket like (\ref{4}) on the surface (\ref{1}).

After this brief review on the Dirac bracket for reducible second-class
constraints, we pass to the implementation of our irreducible procedure. The
solution to the equation (\ref{7}) has the form 
\begin{equation}
d_{\;\;\alpha _{0}}^{\alpha _{1}}=\left( \delta _{\;\;\beta _{1}}^{\alpha
_{1}}+m_{\;\;\;\beta _{1}}^{\alpha _{1}\beta _{0}}\chi _{\beta _{0}}\right) 
\bar{D}_{\;\;\gamma _{1}}^{\beta _{1}}A_{\alpha _{0}}^{\;\;\gamma _{1}},
\label{11}
\end{equation}
where $A_{\alpha _{0}}^{\;\;\gamma _{1}}$ are some functions chosen such
that 
\begin{equation}
rank\left( D_{\;\;\alpha _{1}}^{\gamma _{1}}\right) \equiv rank\left(
Z_{\;\;\alpha _{1}}^{\alpha _{0}}A_{\alpha _{0}}^{\;\;\gamma _{1}}\right)
=M_{1},  \label{12}
\end{equation}
$\bar{D}_{\;\;\gamma _{1}}^{\beta _{1}}$ stands for the inverse of $%
D_{\;\;\alpha _{1}}^{\gamma _{1}}$, and $m_{\;\;\;\beta _{1}}^{\alpha
_{1}\beta _{0}}$ are some arbitrary functions. Inserting (\ref{11}) in (\ref
{8}) we find 
\begin{equation}
M^{\alpha _{0}\beta _{0}}C_{\beta _{0}\gamma _{0}}\approx D_{\;\;\gamma
_{0}}^{\alpha _{0}},  \label{13}
\end{equation}
with 
\begin{equation}
D_{\;\;\gamma _{0}}^{\alpha _{0}}=\delta _{\;\;\gamma _{0}}^{\alpha
_{0}}-Z_{\;\;\alpha _{1}}^{\alpha _{0}}\bar{D}_{\;\;\gamma _{1}}^{\alpha
_{1}}A_{\gamma _{0}}^{\;\;\gamma _{1}}.  \label{14}
\end{equation}
With these elements at hand, the next theorem is shown to hold.

\begin{theorem}
There exists an invertible antisymmetric matrix $\mu ^{\alpha _{0}\beta _{0}}
$ such that the Dirac bracket (\ref{10}) takes the form 
\begin{equation}
\left[ F,G\right] ^{*}=\left[ F,G\right] -\left[ F,\chi _{\alpha
_{0}}\right] \mu ^{\alpha _{0}\beta _{0}}\left[ \chi _{\beta _{0}},G\right] ,
\label{24}
\end{equation}
on the surface (\ref{1}).
\end{theorem}

\proof%
First, we observe that the matrix (\ref{14}) is a projector 
\begin{equation}
D_{\;\;\gamma _{0}}^{\alpha _{0}}D_{\;\;\lambda _{0}}^{\gamma
_{0}}=D_{\;\;\lambda _{0}}^{\alpha _{0}},  \label{15}
\end{equation}
and satisfies the relations 
\begin{equation}
D_{\;\;\gamma _{0}}^{\alpha _{0}}Z_{\;\;\gamma _{1}}^{\gamma _{0}}=0,
\label{16}
\end{equation}
\begin{equation}
A_{\alpha _{0}}^{\;\;\gamma _{1}}D_{\;\;\gamma _{0}}^{\alpha _{0}}=0,
\label{17}
\end{equation}
\begin{equation}
D_{\;\;\gamma _{0}}^{\alpha _{0}}\chi _{\alpha _{0}}=\chi _{\gamma _{0}}.
\label{18}
\end{equation}
Multiplying (\ref{13}) by $A_{\alpha _{0}}^{\;\;\gamma _{1}}$ and using (\ref
{17}) we obtain the relations $A_{\alpha _{0}}^{\;\;\gamma _{1}}M^{\alpha
_{0}\beta _{0}}C_{\beta _{0}\gamma _{0}}\approx 0$, which then lead to 
\begin{equation}
A_{\alpha _{0}}^{\;\;\gamma _{1}}M^{\alpha _{0}\beta _{0}}\approx 0.
\label{19}
\end{equation}
The relations (\ref{19}) allow us to represent $M^{\alpha _{0}\beta _{0}}$
under the form 
\begin{equation}
M^{\alpha _{0}\beta _{0}}\approx D_{\;\;\lambda _{0}}^{\alpha _{0}}\mu
^{\lambda _{0}\sigma _{0}}D_{\;\;\sigma _{0}}^{\beta _{0}},  \label{20}
\end{equation}
where $\mu ^{\lambda _{0}\sigma _{0}}$ is an antisymmetric matrix. Now, we
prove that $\mu ^{\lambda _{0}\sigma _{0}}$ is invertible. On account of (%
\ref{16}), the solution to (\ref{20}) reads as 
\begin{equation}
\mu ^{\lambda _{0}\sigma _{0}}\approx M^{\lambda _{0}\sigma
_{0}}+Z_{\;\;\lambda _{1}}^{\lambda _{0}}\bar{D}_{\;\;\beta _{1}}^{\lambda
_{1}}\omega ^{\beta _{1}\gamma _{1}}\bar{D}_{\;\;\gamma _{1}}^{\sigma
_{1}}Z_{\;\;\sigma _{1}}^{\sigma _{0}},  \label{21}
\end{equation}
for an invertible antisymmetric matrix $\omega ^{\beta _{1}\gamma _{1}}$. As
the only null vectors of $M^{\alpha _{0}\beta _{0}}$ are $A_{\alpha
_{0}}^{\;\;\gamma _{1}}$ (see (\ref{19})), it results that $A_{\lambda
_{0}}^{\;\;\rho _{1}}\mu ^{\lambda _{0}\sigma _{0}}\approx \omega ^{\rho
_{1}\gamma _{1}}\bar{D}_{\;\;\gamma _{1}}^{\sigma _{1}}Z_{\;\;\sigma
_{1}}^{\sigma _{0}}$ vanish if and only if $Z_{\;\;\sigma _{1}}^{\sigma
_{0}} $ vanish (because $\omega ^{\rho _{1}\gamma _{1}}\bar{D}_{\;\;\gamma
_{1}}^{\sigma _{1}}$ is invertible). However, by assumption we have that not
all $Z_{\;\;\sigma _{1}}^{\sigma _{0}}$ vanish, so it results that $\omega
^{\rho _{1}\gamma _{1}}\bar{D}_{\;\;\gamma _{1}}^{\sigma _{1}}Z_{\;\;\sigma
_{1}}^{\sigma _{0}}$ is nonvanishing. In consequence, $\mu ^{\lambda
_{0}\sigma _{0}}$ has no null vectors, being therefore invertible. Inserting
(\ref{20}) in (\ref{10}) and using (\ref{18}), we deduce precisely (\ref{24}%
). This proves the theorem. $\Box $

Formulas (\ref{13}) and (\ref{20}) allow us to represent $C_{\beta
_{0}\gamma _{0}}$ like 
\begin{equation}
C_{\beta _{0}\gamma _{0}}\approx D_{\;\;\beta _{0}}^{\rho _{0}}\mu _{\rho
_{0}\tau _{0}}D_{\;\;\gamma _{0}}^{\tau _{0}},  \label{22}
\end{equation}
which gives 
\begin{equation}
\mu _{\rho _{0}\tau _{0}}\approx C_{\rho _{0}\tau _{0}}+A_{\rho
_{0}}^{\;\;\rho _{1}}\omega _{\rho _{1}\tau _{1}}A_{\tau _{0}}^{\;\;\tau
_{1}},  \label{23}
\end{equation}
where $\mu _{\rho _{0}\tau _{0}}$ and $\omega _{\rho _{1}\tau _{1}}$ stand
for the inverses of the corresponding upper-indices matrices. It is easy to
see that (\ref{20}) and (\ref{22}) verify (\ref{13}). Apart from being
antisymmetric and invertible, the matrix $\omega _{\rho _{1}\tau _{1}}$ is
up to our choice. In order to endow this matrix with a concrete
significance, we introduce some new variables $\left( y_{\alpha _{1}}\right)
_{\alpha _{1}=1,\cdots ,M_{1}}$ with the Poisson brackets 
\begin{equation}
\left[ y_{\alpha _{1}},y_{\beta _{1}}\right] =\omega _{\alpha _{1}\beta
_{1}},  \label{25}
\end{equation}
and consider the system subject to the reducible second-class constraints 
\begin{equation}
\chi _{\alpha _{0}}\approx 0,\;y_{\alpha _{1}}\approx 0.  \label{26}
\end{equation}
The Dirac bracket on the phase-space described by $\left( z^{a},y_{\alpha
_{1}}\right) $ corresponding to the above second-class constraints reads as 
\begin{equation}
\left. \left[ F,G\right] ^{*}\right| _{z,y}=\left[ F,G\right] -\left[ F,\chi
_{\alpha _{0}}\right] \mu ^{\alpha _{0}\beta _{0}}\left[ \chi _{\beta
_{0}},G\right] -\left[ F,y_{\alpha _{1}}\right] \omega ^{\alpha _{1}\beta
_{1}}\left[ y_{\beta _{1}},G\right] ,  \label{27}
\end{equation}
where the Poisson brackets from the right hand-side of (\ref{27}) contain
derivatives with respect to all $z^{a}$ and $y_{\alpha _{1}}$. After some
computation we infer that 
\begin{equation}
\left. \left[ F,G\right] ^{*}\right| _{z,y}\approx \left[ F,G\right] ^{*},
\label{28}
\end{equation}
where $\left[ F,G\right] ^{*}$ is given by (\ref{24}) and the weak equality
refers to the surface (\ref{26}). Under these considerations, we are able to
prove the following theorem.

\begin{theorem}
(i) There exist some irreducible second-class constraints equivalent to (\ref
{26}) 
\begin{equation}
\tilde{\chi}_{\alpha _{0}}\left( z^{a},y_{\alpha _{1}}\right) \approx 0,
\label{29}
\end{equation}
such that 
\begin{equation}
\left[ \tilde{\chi}_{\alpha _{0}},\tilde{\chi}_{\beta _{0}}\right] \approx
\mu _{\alpha _{0}\beta _{0}}.  \label{30}
\end{equation}

(ii) The Dirac bracket with respect to the irreducible second-class
constraints (\ref{29}) 
\begin{equation}
\left. \left[ F,G\right] ^{*}\right| _{{\rm ired}}=\left[ F,G\right] -\left[
F,\tilde{\chi}_{\alpha _{0}}\right] \mu ^{\alpha _{0}\beta _{0}}\left[ 
\tilde{\chi}_{\beta _{0}},G\right] ,  \label{31}
\end{equation}
coincides with (\ref{27}) 
\begin{equation}
\left. \left[ F,G\right] ^{*}\right| _{{\rm ired}}=\left. \left[ F,G\right]
^{*}\right| _{z,y},  \label{32}
\end{equation}
on the surface (\ref{29}).
\end{theorem}

\proof%
(i) Using (\ref{23}), formulas (\ref{30}) become 
\begin{equation}
\left[ \tilde{\chi}_{\alpha _{0}},\tilde{\chi}_{\beta _{0}}\right] \approx
C_{\alpha _{0}\beta _{0}}+A_{\alpha _{0}}^{\;\;\rho _{1}}\omega _{\rho
_{1}\tau _{1}}A_{\beta _{0}}^{\;\;\tau _{1}}.  \label{34}
\end{equation}
Now, we prove that the solution to the above equations is expressed by 
\begin{equation}
\tilde{\chi}_{\alpha _{0}}=\chi _{\alpha _{0}}+A_{\alpha _{0}}^{\;\;\alpha
_{1}}y_{\alpha _{1}}.  \label{35}
\end{equation}
The functions $\tilde{\chi}_{\alpha _{0}}$ are irreducible. Indeed, $%
Z_{\;\;\beta _{1}}^{\alpha _{0}}\tilde{\chi}_{\alpha _{0}}=D_{\;\;\beta
_{1}}^{\alpha _{1}}y_{\alpha _{1}}$ vanish if and only if $y_{\alpha _{1}}$
vanish, so if and only if the new constraints reduce to (\ref{26}). This
proves the irreducibility. After some simple computation, from (\ref{35}) we
infer that 
\begin{equation}
\chi _{\alpha _{0}}=D_{\;\;\alpha _{0}}^{\beta _{0}}\tilde{\chi}_{\beta
_{0}},\;y_{\alpha _{1}}=\bar{D}_{\;\;\alpha _{1}}^{\beta _{1}}Z_{\;\;\beta
_{1}}^{\beta _{0}}\tilde{\chi}_{\beta _{0}}.  \label{36}
\end{equation}
It is easy to see that if (\ref{26}) hold, then (\ref{29}) also hold (with $%
\tilde{\chi}_{\alpha _{0}}$ given by (\ref{35})). From (\ref{36}) we obtain
that if (\ref{29}) hold, (\ref{26}) hold, too, so 
\begin{equation}
\tilde{\chi}_{\alpha _{0}}\approx 0\Leftrightarrow \chi _{\alpha
_{0}}\approx 0,\;y_{\alpha _{1}}\approx 0,  \label{37}
\end{equation}
such that the constraints (\ref{29}) are equivalent to (\ref{26}). Finally,
if we use (\ref{37}), then the functions (\ref{35}) satisfy (\ref{34}). This
proves (i).

(ii) By direct computation we get 
\begin{eqnarray}\label{38}
& &\left. \left[ F,G\right] ^{*}\right| _{{\rm ired}}\approx \left[ F,G\right]
^{*}-\left[ F,\chi _{\alpha _{0}}\right] \mu ^{\alpha _{0}\beta
_{0}}A_{\beta _{0}}^{\;\;\beta _{1}}\left[ y_{\beta _{1}},G\right] - 
\nonumber \\
& &\left[ F,y_{\alpha _{1}}\right] A_{\alpha _{0}}^{\;\;\alpha _{1}}\mu
^{\alpha _{0}\beta _{0}}\left[ \chi _{\beta _{0}},G\right] -\left[
F,y_{\alpha _{1}}\right] A_{\alpha _{0}}^{\;\;\alpha _{1}}\mu ^{\alpha
_{0}\beta _{0}}A_{\beta _{0}}^{\;\;\beta _{1}}\left[ y_{\beta _{1}},G\right]
.  
\end{eqnarray}
On the other hand, from (\ref{21}) we obtain that 
\begin{equation}
\mu ^{\alpha _{0}\beta _{0}}A_{\beta _{0}}^{\;\;\beta _{1}}\approx
Z_{\;\;\alpha _{1}}^{\alpha _{0}}\bar{D}_{\;\;\gamma _{1}}^{\alpha
_{1}}\omega ^{\gamma _{1}\beta _{1}},\;A_{\alpha _{0}}^{\;\;\alpha _{1}}\mu
^{\alpha _{0}\beta _{0}}\approx \omega ^{\alpha _{1}\gamma _{1}}\bar{D}%
_{\;\;\gamma _{1}}^{\beta _{1}}Z_{\;\;\beta _{1}}^{\beta _{0}},  \label{39}
\end{equation}
\begin{equation}
A_{\alpha _{0}}^{\;\;\alpha _{1}}\mu ^{\alpha _{0}\beta _{0}}A_{\beta
_{0}}^{\;\;\beta _{1}}\approx \omega ^{\alpha _{1}\beta _{1}}.  \label{40}
\end{equation}
Inserting (\ref{39}--\ref{40}) in (\ref{38}), we immediately find (\ref{32}%
). This proves (ii).$\Box $

The last theorem proves that we can approach reducible second-class
constraints in an irreducible fashion. Thus, starting with the reducible
constraints (\ref{1}) we construct the irreducible constraint functions (\ref
{35}), whose Poisson brackets form an invertible matrix. Formulas (\ref{28})
and (\ref{32}) ensure that $\left. \left[ F,G\right] ^{*}\right| _{{\rm ired}%
}\approx \left[ F,G\right] ^{*}$, so the fundamental Dirac brackets among
the original variables $z^{a}$ within the irreducible setting coincide with
those from the reducible version 
\begin{equation}
\left. \left[ z^{a},z^{b}\right] ^{*}\right| _{{\rm ired}}\approx \left[
z^{a},z^{b}\right] ^{*}.  \label{41}
\end{equation}
Moreover, the new variables $y_{\alpha _{1}}$ do not affect the irreducible
Dirac bracket as from (\ref{32}) we have that $\left. \left[ y_{\alpha
_{1}},F\right] ^{*}\right| _{{\rm ired}}\approx 0$. Thus, the equations of
motion for the original reducible system can be written as $\dot{z}%
^{a}\approx \left. \left[ z^{a},H\right] ^{*}\right| _{{\rm ired}}$, where $%
H $ is the canonical Hamiltonian. The equations of motion for $y_{\alpha
_{1}}$ read as $\dot{y}_{\alpha _{1}}\approx 0$, and lead to $y_{\alpha
_{1}}=0$ by taking some appropriate boundary conditions (vacuum to vacuum)
for these unphysical variables. This completes the general procedure.

Let us briefly exemplify the general theory on gauge-fixed two-forms,
subject to the second-class constraints 
\begin{equation}
\chi _{\alpha _{0}}\equiv \left( 
\begin{array}{c}
-2\partial ^{k}\pi _{ki} \\ 
-\partial _{l}A^{lj}
\end{array}
\right) \approx 0.  \label{42}
\end{equation}
The constraints involving the temporal components of the two-form and its
momenta are irreducible, and will be omitted. The constraints (\ref{42}) are
first-stage reducible, with the reducibility functions expressed by 
\begin{equation}
Z_{\;\;\alpha _{1}}^{\alpha _{0}}=\left( 
\begin{array}{cc}
\partial ^{i} & 0 \\ 
0 & \partial _{j}
\end{array}
\right) .  \label{43}
\end{equation}
Acting along the line exposed in the above, we take the matrix $A_{\alpha
_{0}}^{\;\;\alpha _{1}}$ under the form 
\begin{equation}
A_{\alpha _{0}}^{\;\;\alpha _{1}}=\left( 
\begin{array}{cc}
-\partial _{i} & 0 \\ 
0 & -\partial ^{j}
\end{array}
\right) ,  \label{44}
\end{equation}
so 
\begin{equation}
D_{\;\;\beta _{1}}^{\alpha _{1}}=\left( 
\begin{array}{cc}
-\partial _{i}\partial ^{i} & 0 \\ 
0 & -\partial ^{j}\partial _{j}
\end{array}
\right) ,  \label{45}
\end{equation}
is invertible. In order to construct the irreducible second-class
constraints, we introduce the variables 
\begin{equation}
y_{\alpha _{1}}=\left( 
\begin{array}{c}
\pi  \\ 
\varphi 
\end{array}
\right) ,  \label{46}
\end{equation}
and take 
\begin{equation}
\omega _{\alpha _{1}\beta _{1}}=\left( 
\begin{array}{cc}
0 & -1 \\ 
1 & 0
\end{array}
\right) .  \label{47}
\end{equation}
As can be seen, the supplementary scalar fields $\left( \pi ,\varphi \right) 
$ are canonically conjugated, with $\pi $ the momentum. Then, the
irreducible second-class constraints are expressed by 
\begin{equation}
\tilde{\chi}_{\alpha _{0}}\equiv \left( 
\begin{array}{c}
-2\partial ^{k}\pi _{ki}-\partial _{i}\pi  \\ 
-\partial _{l}A^{lj}-\partial ^{j}\varphi 
\end{array}
\right) \approx 0,  \label{48}
\end{equation}
such that 
\begin{equation}
\mu _{\alpha _{0}\beta _{0}}=\left( 
\begin{array}{cc}
0 & \delta _{i}^{\;\;j}\triangle  \\ 
-\delta _{\;\;l}^{k}\triangle  & 0
\end{array}
\right) ,  \label{49}
\end{equation}
where $\triangle =\partial ^{l}\partial _{l}$. By inverting (\ref{49}) we
obtain that the only nonvanishing irreducible Dirac brackets are given by 
\begin{equation}
\left. \left[ A^{ij}\left( x\right) ,\pi _{kl}\left( y\right) \right]
^{*}\right| _{{\rm ired}}=\frac{1}{2}\left( \delta _{\;\;k}^{\left[ i\right.
}\delta _{l}^{\left. j\right] }+\frac{1}{\triangle }\partial ^{\left[
i\right. }\delta _{\;\;p}^{\left. j\right] }\partial _{\left[ k\right.
}\delta _{\;\;\left. l\right] }^{p}\right) \delta ^{3}\left( {\bf x}-{\bf y}%
\right) ,  \label{50}
\end{equation}
where the notation $\left[ i_{1}\cdots i_{n}\right] $ means antisymmetry
with respect to the indices between brackets. The result given by (\ref{50})
reproduces the standard result from the literature \cite{8,9}. By
means of (\ref{50}) and of the canonical Hamiltonian associated with the
model under study we can immediately write down the corresponding equations
of motion. This completes the analysis of the example.

To conclude with, in this paper we have exposed an irreducible procedure for
approaching systems with reducible second-class constraints. Our strategy
includes three main steps. First, we express the Dirac bracket for the
reducible system in terms of an invertible matrix. Second, we establish the
equality between this Dirac bracket and that corresponding to the
intermediate theory, based on the constraints (\ref{26}). Third, we prove
that there exists an irreducible second-class constraint set equivalent with
(\ref{26}) such that the corresponding Dirac brackets coincide. These three
steps enforce the fact that the fundamental Dirac brackets with respect to
the original variables derived within the irreducible and original reducible
settings coincide. Moreover, the newly added variables do not affect the
Dirac bracket, so the canonical approach to the initial reducible system can
be developed in terms of the Dirac bracket corresponding to the irreducible
theory. Finally, the general procedure was exemplified for gauge-fixed
two-forms.

\section*{Acknowledgment}

Two of the authors (C.B. and S.O.S.) acknowledge financial support from a 
Romanian National Council for Academic Scientific Research (CNCSIS) grant.

\end{document}